\documentclass{article}
\usepackage[utf8]{inputenc}
\usepackage{charter} 
\usepackage{geometry}
\usepackage{authblk}
\usepackage{siunitx}
\usepackage{amsfonts}
\usepackage{hyperref}
\usepackage{graphicx, caption}
\usepackage{xcolor}
\usepackage[style=nature]{biblatex}
\usepackage{adjustbox}
\usepackage{tabularx}
\usepackage{mathtools}
\usepackage[version=3]{mhchem} 
\usepackage{amssymb}
\usepackage{braket}
\usepackage{esvect}
\usepackage{listings}
\usepackage{mathptmx}
\usepackage{amsmath}
\usepackage{gensymb}

\DeclareUnicodeCharacter{0300}{*************************************}

\graphicspath{ {./figures/} }

\title{Hydrogen in Disordered Titania:
Connecting Local Chemistry, Structure, and Stoichiometry through Accelerated Exploration }

\author[1,4]{James Chapman \thanks{Corresponding Author, jc112358@bu.edu}}
\author[1]{Kyoung E. Kweon \thanks{Corresponding Author, kweon1@llnl.gov}}
\author[1]{Yakun Zhu}
\author[1,3]{Kyle Bushick}
\author[1]{Leonardus Bimo Bayu Aji}
\author[1]{Christopher A. Colla}
\author[1,2]{Nir Goldman}
\author[1]{Nathan Keilbart}
\author[1]{S. Roger Qiu}
\author[1]{Tae Wook Heo}
\author[1]{Brandon C. Wood\thanks{Corresponding Author, wood37@llnl.gov}}

\affil[1]{Materials Science Division, Lawrence Livermore National Laboratory, Livermore, CA, USA}
\affil[2]{Department of Chemical Engineering, University of California, Davis, California 95616, United States}
\affil[3]{Department of Materials Science and Engineering, University of Michigan, Ann Arbor, Michigan 48109, United States}
\affil[4]{Department of Mechanical Engineering, Boston University, Boston, Massachusetts 02215, United States}

\begin{document}

\maketitle

\begin{abstract}

Hydrogen incorporation in native surface oxides of metal alloys often controls the onset of metal hydriding, with implications for materials corrosion and hydrogen storage. A key representative example is titania, which forms as a passivating layer on a variety of titanium alloys for structural and functional applications. These oxides tend to be structurally diverse, featuring polymorphic phases, grain boundaries, and amorphous regions that generate a disparate set of unique local environments for hydrogen. Here, we introduce a workflow that can efficiently and accurately navigate this complexity. First, a machine learning force field, trained on ab initio molecular dynamics simulations, was used to generate amorphous configurations. Density functional theory calculations were then performed on these structures to identify local oxygen environments, which were compared against experimental observations. Second, to classify subtle differences across the disordered configuration space, we employ a graph-based sampling procedure. Finally, local hydrogen binding energies are computed using exhaustive density functional theory calculations on representative configurations. We leverage this methodology to show that hydrogen binding energetics are described by local oxygen coordination, which in turn is affected by stoichiometry. Together these results imply that hydrogen incorporation and transport in TiO$_x$ can be tailored through compositional engineering, with implications for improving performance and durability of titanium-derived alloys in hydrogen environments.

\end{abstract}
\section{Introduction}


Ultimately, corrosion is a thermodynamically driven process that affects all metal/metal-oxide systems and alloys, usually through one of several mechanisms, such as hydrogen embrittlement, galvanic processes, or oxidation \cite{vachtsevanos_natarajan_rajamanai_sandborn_2020}. Hydrogen embrittlement in particular can cause severe and dangerous damage due to the spontaneous formation of metal hydrides. These reactions are sometimes pyrophoric, and the hydrides themselves are frequently dispersive powders that are highly toxic \cite{GanyNetzer_1985_157_168}. As a result, mitigation of this process has ramifications for a number of manufacturing efforts, including actinides processing \cite{ewing_runde_albrecht-schmitt_2010}, stainless steal reinforcement, etc.  

Typically, transition metals such as titanium are naturally covered with a protective oxide layer than inhibits embrittlement\cite{schmuki_2002}. This protection is not permanent though, with chemical species such as oxygen and hydrogen eventually diffusing through the oxide and into the bulk metal/alloy \cite{kofstad_1995,doi:10.1063/1.2768951}. This sequence of events implies that the incubation (initiation) time, and ultimately failure, of the system is rate-limited by the ability of the corrosive chemical agents to permeate the oxide layer. Hydriding initiation thus most likely begins at this oxide surface, where hydrogen can diffuse through and attack the metal underbelly, thereby starting the conversion of titanium into TiH$_2$.  However, there is still a question of how hydrogen is transported from the surface into the metal itself. The oxide layer is likely polycrystalline \cite{A802619J}, with a number of unique and complex atomic environments existing between the oxide surface and the underlying metal, including grain boundaries, nano/microscale defects such as voids and cracks, and the interface region between the oxide and metal/alloy \cite{henrich_kurtz_1981,doi:10.1063/1.3122984,lu_kaxiras_2005,doi:10.1063/1.109101,WANG201622214}.. There is speculation that thermodynamic driving forces exist that result in larger hydrogen concentration in the grain boundaries, which themselves are likely disordered and resemble amorphous phases\cite{KEBLINSKI1997987}. These then become channels for mass transport to the metallic material underneath.  

As a result, the lifetime of a given metal or alloy could be extended through control over the kinetics of key underlying hydrogen transport processes within the oxide. While the failure of materials under steady-state conditions can be predicted exceptionally well through the use of parameterized empirical models, the precursors to hydriding initiation are still poorly understood \cite{10.1115/1.3063646,SONG20092657,KAMRUNNAHAR2010669,ELMAADDAWY2007168}. This initial stage of the embrittlement process is non-trivial, as its effects are likely coupled together with other complex phenomena (i.e., nucleation and growth of the metal hydride) in possibly complex and unintuitive ways. As no single methodology currently exists to explicitly explore diffusion through all aforementioned features within the oxide layer under dynamic conditions, a piece-meal approach must be taken to understand how hydrogen behaves within each structural environment \cite{YASHIMA20153,ANDERSSON20101461,Sun2010}. 

In this work we investigate the thermodynamics of hydrogen binding within the grain boundary regions of TiO$_{x}$ using a combination of density functional theory (DFT), a machine learning force field, and a graph theoretical structure characterization technique. As previously stated, amorphous titania was chosen as a surrogate for grain boundaries. Explicitly simulating hydrogen diffusion within the amorphous phase under dynamic conditions is non-trivial, as the size and time requirements are outside of the realm of DFT, and classical models are difficult to adequately capture the electronic effects that determine hydrogen binding \cite{C5NJ02836A,doi:10.1021/cr200217c}. 

Therefore, we adopt a three-step computational workflow to overcome these challenges: (1) machine learning force field classical molecular dynamics simulations to explore vast regions of the amorphous phase space, (2) graph-theory driven structure characterization to identify and down-select representative structures, and (3) density functional theory calculations on the sampled configurations, with hydrogen inserted in the system, to obtain the hydrogen binding energy with a high level of fidelity. This pipeline allows for the ``best-of-both-worlds'', where fast and efficient molecular dynamics (MD) simulations can be performed to generate a practically unlimited number of configurations and expensive but accurate DFT calculations can provide a reliable estimation to the spectrum of possible hydrogen binding energies present with the amorphous phase space.

The rest of the paper is as follows. We begin by providing a detailed understanding of the various methodological aspects of our computational and experimental procedures. We then provide a detailed description of the atomic structure of amorphous titania using both experiments and simulations. We show that our simulation framework, which combines nuclear magnetic resonance (NMR), MD, and graph theory, can accurately link the characterization of local oxygen environments to experimental NMR observations. Our MD simulations are also used to understand the effect of oxygen concentration on the likelihood of finding specific oxygen coordination environments within an amorphous sample, and use this information to validate the experimental characterization. We then give a detailed description of the DFT calculated hydrogen binding energies for both stoichiometric and non-stoichiometric amorphous titania. Finally, we discuss how the observations gathered from this work give insight into how structural features such as the oxygen concentration could ultimately be used to tailor properties such as hydrogen diffusion and permeation.

\section{Methods}
\subsection{Computational Methods}
\subsubsection{Computational workflow}

\begin{figure}
       \centering
    	\includegraphics[trim={0 0cm 0 0cm},width=0.8\textwidth]{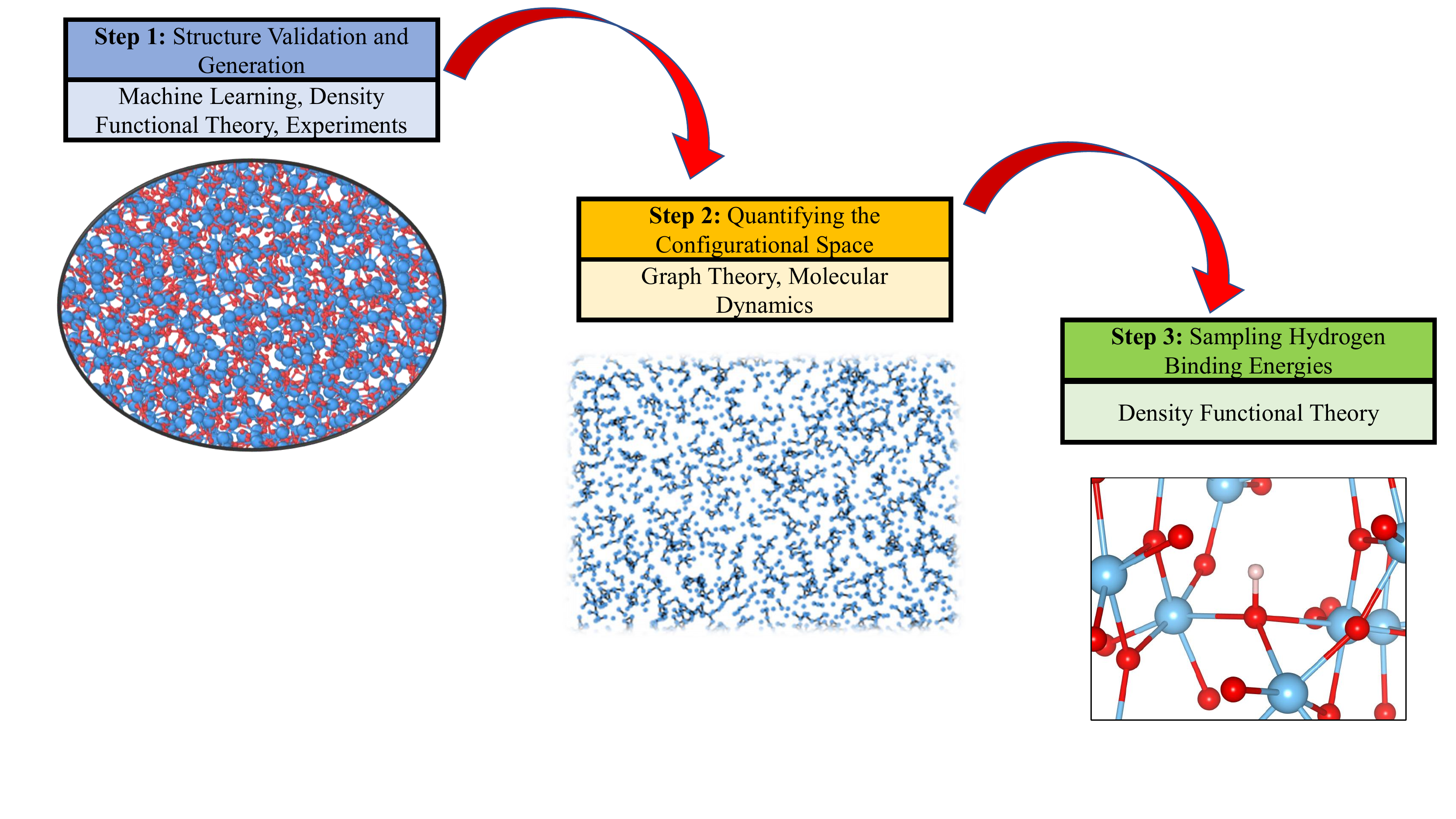}
    	\caption{Step 1: the atomic structure of amorphous titania is generated both via ab initio and machine learning-driven MD simulations and experimental synthesis. This atomic structure is then validated through the use of NMR experiments and simulations. Step 2: bounds on the amorphous phase space are quantified by employing a graph characterization technique on the machine learning-driven MD simulations, providing an accurate description of the amorphous configurational space. Step 3: hydrogen binding energies are calculated using DFT in a high throughput manner.}
    \label{fig:workflow}
\end{figure}

The computational workflow used in this work employs a 3-step approach, in which long time-scale MD simulations are performed with an atomic force neural network (AFNN) and then a characterized, geometrically, with a graph-based order parameter. Representative structures are then down-selected using a combination of K-means clustering and stochastic random sampling. DFT calculations are then performed to obtain the hydrogen binding energy of the down-selected structures. This workflow combines the advantages of each methodology: (1) machine learning force fields can explore vast regions of the configurational space not attainable with DFT, (2) graph theory provides a physics-informed phase space characterization, making sampling intuitive and efficient, and (3) high-fidelity DFT binding energy calculation ensures that the resulting energy distributions are accurate and reliable. Fig. \ref{fig:workflow} provides a visual depiction of this computational workflow. 

\subsubsection{Electronic structure details}

DFT calculations were performed using the VASP package \cite{KRESSE199615} with projector augmented wave (PAW) pseudo potentials \cite{PhysRevB.50.17953}.  As valence electrons, we considered the 3p$^{6}$3d$^{2}$4s$^{2}$ electron configuration for Ti and 2s$^{2}$2p$^{4}$ for O, respectively.  The PBE exchange-correlation functional revised for solids (PBEsol) \cite{PhysRevLett.100.136406} was used to compute structural properties of bulk titania more accurately \cite{C6CP04973G}.  

\subsubsection{Ab initio molecular dynamics}

To generate amorphous structures, the melt-quench method was applied for TiO$_{2}$ (having 72 Ti and 144 O atoms) and TiO$_{1.88}$ (having 72 Ti and 135 O atoms) systems using AIMD simulations \cite{doi:10.1021/acs.jpclett.8b01067,doi:10.1063/1.5042783,doi:10.1063/1.4998611}.  Starting from 72 Ti and 144 O atoms in a cubic cell with two different crystalline atomic configurations (rutile and anatase), the randomized TiO$_{2}$ structures were generated by very high temperature AIMD at 5000 K in the canonical (NVT) ensemble for 10 ps with a 2 fs time step; similarly, two randomized TiO$_{1.88}$ structures were created from crystalline Ti$_{72}$O$_{144}$ by removing 9 O atoms.  The randomized structures are cooled to the liquid temperature of 2250 K during 2 ps and subsequently equilibrated at 2250 K for 5 ps in NVT.  After that, two snapshots were taken every 5 ps from further dynamics at 2250 K within the microcanonical (NVE) ensemble to improve statistical distribution of local and independent melt structures.  Note that the experimental density of 3.21 g/cm$^{3}$ for the liquid TiO$_{2}$ \cite{https://doi.org/10.1111/j.1151-2916.1991.tb06833.x} was used for the simulated liquid model at 2250 K (the same density of 3.21g/cm$^{3}$ is used for the TiO$_{1.88}$ at 2250 K).  

The final structures were obtained by taking each melted snapshot and quenching to 300 K with a cooling rate of 19.5 K/ps, for an overall quenching time of 100 ps.  Here, the quench process was performed using the NVT ensemble, while the density of system was rescaled every 8 ps (equivalent to 156 K) to account realistic density change of the liquid \cite{https://doi.org/10.1111/j.1151-2916.1991.tb06833.x,PhysRevB.90.094204} during the quenching as suggested by Mavra\v{c}i\'{c} et. al \cite{doi:10.1021/acs.jpclett.8b01067}.  Finally, the quenched structures were equilibrated at 300 K for 10 ps under NVT conditions, using DFT+U calculations (as will be discussed later), to ensure that the structure is fully optimized under ambient pressure. For each TiO$_{2}$ and TiO$_{1.88}$, four different independent structures were obtained, while the density at 300K is further optimized, resulting in the average final densities of 3.72 g/cm$^{3}$ for TiO$_{2}$ and 3.55 g/cm$^{3}$ for TiO$_{1.88}$, respectively; note that with energy variations within the structures, only three structures with lower energies were taken for additional calculations.  For all AIMD simulations, a time step of 1 fs was employed unless mentioned otherwise. For all NVT simulations, the Nos\'{e}-Hoover thermostat was used. 

\subsubsection{Ab initio NMR details}

The isotropic chemical shift ($\delta_{iso}$) for O sites is computed as $\delta_{iso}$ = $\delta_{cal}$ + $\delta_{ref}$, where $\delta_{cal}$ is the calculated chemical shift via the linear response method in VASP and $\delta_{ref}$  is the chemical shift for the reference compound \cite{C6CC02542K,doi:10.1126/sciadv.1400133,Li2017,C4SC00083H}.  We included the contribution from the core electrons as well as the valence electrons to obtain $\delta_{cal}$.  In this work, $\delta_{ref}$ is determined by aligning $\delta_{cal}$ for O in anatase to the corresponding experimental $\delta_{iso}$ (557 ppm ) \cite{JM9930300697}.  To accurately calculate the variation in the electronic structures within the linear response approach, higher energy cutoff of 600 eV and denser 2x2x2 Monkhorst-Pack k-grid were used for the 96 atom supercell.  With ~400 different amorphous models, a total of ~30,000  $\delta_{iso}$ datapoints for O sites were collected, which is large enough to capture the correlation between $\delta_{iso}$ for O sites and their coordination environment.

\subsubsection{Ab initio structure optimization}
After the melt-quenching AIMD simulations, structural optimization for the amorphous TiO$_{x}$ ($x$ = 2 or 1.88) was carried out using spin-polarized DFT calculations to include the electron spin polarization with the plane-wave energy cutoff of 450 eV. We also employed DFT+U within Dudarev’s approach \cite{PhysRevB.57.1505} to correct the electronic self-interaction error for the localized 3d states, which becomes crucial to accurately describe structural and electronic properties of oxygen deficient titania with H interstitials and/or O vacancies \cite{doi:10.1063/1.2996362,doi:10.1063/1.3617244,doi:10.1021/acs.jpcc.5b05338}; a value of Hubbard U = 4 eV is applied for all Ti 3d states \cite{Li2017}.  The Brillouin zone integration is approximated using a single k-point ($\Gamma$-point), which could be sufficient for the reasonably large supercells (containing 72 TiO$_{x}$ formula units) of insulating systems with a relatively small computational cost.  All structures were relaxed until the force on each atom is less than 0.02 eV/$\AA$.

\subsubsection{Atomic force neural network details}

In this work, atomic forces are learned by establishing a mapping between the atomic features and their respective atomic force components using a deep neural network (NN). The AFNN framework was designed based on previous works regarding machine learning force matching schemes \cite{Chapman2020,CHAPMAN2020109483,https://doi.org/10.1002}. As there are multiple species present, two AFNNs are employed to independently predict the atomic forces acting on a given chemical element: one AFNN for Ti, and another for O. A visual description of this arrangement can be found in the supplemental information. The NN architecture of both AFNNs employs an input layer containing a neuron count equal to the number of atomic features, one hidden layer containing a neuron count equal to the input layer, a second hidden layer containing a neuron count equal to four times that of the input layer, and a single output that maps to a given force component. Therefore, each atom will make three independent NN predictions to account for the $x$, $y$, and $z$ atomic force components. Each AFNN employed the $tanh$ activation function throughout all hidden layer neurons, due to its symmetry about $x=0$, which matches the expected symmetry of the atomic forces with respect to the atomic features. A bias vector is also associated with each layer in the AFNN, helping to control the values passed into each neuron's activation function.

During the AFNN model's training phase only TiO$_{2}$ was used to train the model, while the non-stoichiometric DFT data (TiO$_{1.88}$) was used to validate  the model's ability to extrapolate to unseen environments. 75\% of the reference data was used sampled using the K-Means clustering \cite{LIKAS2003451} algorithm within the atomic feature space (using 100 randomly initialized clusters). After the K-means clusters had been optimized, training data was randomly sampled from within each cluster until the desired number of training points had been met. The remaining 25\% split equally into validation and test sets. 10000 epochs were used to ensure convergence in the model's predictions, along with the Adamax optimization algorithm \cite{doi:10.1080/01431161.2019.1694725}. More details regarding the AFNN training can be found in the supplemental information. 

\subsubsection{Classical structure generation details}

All structures used in this work were generated via MD using the AFNN described in the previous sections, using the LAMMPS \cite{lammps} software. A 96-atom crystal structure in the rutile phase was used as the initial configuration. MD was then performed at 4000K to superheat and liquify the rutile crystal. The cell shape was then slowly modified, manually, until the lattice vectors reached a cubic configuration. The system temperature was then brought down to 2250K, a temperature previously reported as showing the existence of the amorphous phase \cite{https://doi.org/10.1111/j.1151-2916.1991.tb06833.x}. The lattice vectors were again altered, isotropically, until the volume-to-atom ratio reached 12.40 $\frac{\AA^{3}}{atom}$, which was obtained from previously calculated values \cite{doi:10.1021/acs.jpclett.8b01067,doi:10.1063/1.5042783,doi:10.1063/1.4998611}. The system was then equilibrated at 2250K using the finalized system volume. A 25ns MD simulation was then performed in the NVT ensemble at 2250K to generate a large dataset of possible amorphous configurations. 

A larger system, containing 1944 atoms was also used to justify the use of the 96-atom configurations by observing that both trajectories live within the same region of phase space, a point that discussed later. The same procedure described above was used to generate the 1944-atom structures. However, due to issues with the AFNN, the same density could not be attained for the 1944-atom system. Large voids opened within the configuration, potentially due to periodic box effects. Therefore, the 1944-atom system was slightly compressed to 11.74 $\frac{\AA^{3}}{atom}$, which was attained by minimizing the differences in the radial distribution function between the 1944-atom and 96-atom trajectories. While the use of the slightly compressed system may leave out potential periodic size effects at the correct density, the matching of the RDF ensures that a direct comparison can still be made between the two trajectories, as properties such as the oxygen coordination number and nearest neighbor distances will remain nearly identical.

For both the 96-atom and 1944-atom trajectories, several non-stoichiometric cases were also considered: (a) TiO$_{1.85}$, TiO$_{1.9}$, TiO$_{1.95}$ for the 1944-atom system, and (b) TiO$_{1.88}$ for the 96-atom system. These trajectories were generated by first taking the perfect amorphous system and removing the corresponding number of oxygen atoms at random. MD was then performed at 2250K in the NVT ensemble with the oxygen deficient system for 25ns. All non-stoichiometric MD trajectories were generated using the same volume as their respective stoichiometric cases.

For the 96-atom configurations hydrogen was also inserted into the TiO$_{1.88}$ and TiO$_{2}$ systems in order to calculate the hydrogen binding energy. Using the down-selected sites described earlier, 5 random oxygen atoms were chosen to form an O-H bond from each configuration such that no two oxygen atoms of the 5 initial sites were closer than 3\AA. In the event that any pair of oxygen atoms were closer than 3\AA, one of the two was ignored. Any duplicate oxygen sites were also removed during this process. Out of all initial configurations, no snapshot has less than 3 possible hydrogen sites, resulting in 6,865 and 7,556 TiO$_{x}$H structures for $x = $ 2 and 1.88 respectively. All TiO$_{x}$H configurations were then relaxed using DFT in order to obtained the final configuration.

\subsubsection{Graph-based atomic structure characterization} 

The diversity and complexity of the amorphous titania phase space necessitates the efficient and reliable characterization of its atomic structures. In this work, we employ a graph-based methodology, Graph Coordination Network (GCN) \cite{chapman2021sgop}, to classify local pairwise atomic environments contained within a configuration of atoms. Fig. SX provides a graphical visualization of how the GCNs are constructed within amorphous titania. These weighted networks encode radial distances as $1/r_{ij}^{2}$, where $i$ and $j$ are the atomic IDs of two atoms. This information is then mapped into a graph order parameter (SGOP), which encodes the connectivity and shape of the graph by leveraging information contained in the set of unique degrees over the graph. Several SGOPs, defined by their radial cutoff distances, are finally grouped together within a single vector, referred to as the Vector Graph Order Parameter (VGOP). Further information regarding the theory behind this methodology can be found in the supplemental information.

\subsection{Experimental Methods}

\subsubsection{Material synthesis}
Titanium isopropoxide (obtained from Aldrich) was mixed with 17O (35-40\%) labeled demineralized water \cite{Borghols_2010} (obtained from Cambridge Isotope Lab, Inc) at a molar ratio of 1:4. The liquid mixture was stirred to accelerate precipitation. The resulting white precipitate was subsequently left to dry in a furnace in air and at 100 °C for 3 days prior to analysis.

\subsubsection{Characterization and analysis}
A Bruker D8 DISCOVER X-ray diffractometer (XRD) was used for phase analysis with a step size of 0.01$\textsuperscript{\textdegree}$/step, a dwell time of 2 s/step, and a scan range of 20 – 80$\textsuperscript{\textdegree}$. Microstructure analysis was carried out using a FEI 80-300 Titan transmission electron microscope (TEM), equipped with a four-detector energy dispersive x-ray spectroscopy (EDS) system known as ChemiSTEMTM. The sample was prepared by spreading powders on a Cu grid. 

Titanium oxide (TiO$_{x}$) films were prepared by pressing the TiO$_{x}$ powders onto planar carbon substrate. Rutherford backscattering spectrometry (RBS) was employed to measure the compositional depth profile in the TiO$_{x}$ films. Samples were bombarded with a 2 MeV 4He+ ion beam incident between 0 and 10 to the sample surface (to minimize ion channeling in textured films) and scattered into a detector at 165$\textsuperscript{\textdegree}$ from the incident beam direction.  The analysis of RBS spectra was performed with the RUMP code \cite{wang_2010,PADAYACHEE2001122}. with the stoichiometry of O/Ti=2±0.2 the simulation yields the best fitting to the measured data.  The best fit to the experimental data is obtained with O/Ti = 2.

XPS was performed on a PHI Quantum 2000 Scanning ESCA Microprobe using a monochromated Al Ka x-rays (1486.6 eV). Calibration was performed using Au 4f$_{7/2}$ at 84.1 eV and the take-off angle was 60$\textsuperscript{\textdegree}$. The X-ray spot size was around 200 µm. Survey spectra for each sample were recorded using a pass energy of 100 eV and a step size of 1 eV \cite{SENBRITAIN2021150433}, and high-resolution spectra of each sample for the C 1s, O1s, Ti 2p, and Ti 3p regions were recorded with a pass energy of 20 eV and a step size of 0.1 eV. The C1s peak at 284.8 eV for adventitious carbon was used as a reference for all spectra \cite{SENBRITAIN2021150433}.  High resolution spectra were curve-fitted using Multipak 9.6.15 using a Shirley background subtraction. Gaussian-Lorentzian peaks were used for curve fitting \cite{SENBRITAIN2021150433}.

Solid-state 17O and 1H NMR experiments were conducted on a Bruker Avance III NMR spectrometer at 600 MHz. Samples were loaded into 2.5mm NMR rotors and spun at 50kHz. Tap water was used as a chemical shift reference for 17O and TMS for 1H. NMR spectra fitting and quantification was performed using the “dmfit” software package \cite{https://doi.org/10.1002/mrc.984}.  

\subsection{Software Tools}
All DFT calculations were performed using the VASP software \cite{KRESSE199615}. All AFNN MD simulations were performed using the LAMMPS software \cite{lammps}. All atomistic visualizations were created using the OVITO software \cite{Stukowski_2009}. All plots were created using the Matplotlib software \cite{Hunter:2007}. Fitted histograms shown in Fig. \ref{fig:Hbind} were created using SciPy \cite{2020SciPy-NMeth}. All PCA anlaysis were performed using the scikit-learn software \cite{scikit-learn}. Experimental NMR spectra fitting and quantification was obtained through the “dmfit” software \cite{https://doi.org/10.1002/mrc.984}. High resolution spectra were curve-fitted using the Multipak software version 9.6.15.

\section{Atomic force neural network}
\subsection{Atomic-level structural descriptors}

An atom's local geometry is decomposed into numerical descriptors which are then mapped to a specific atomic force component. These descriptors aim to capture unique pairwise aspects of an atom's local atomic environment by observing the changes in the atomic probability densities at various intervals around the atom. The functional form of the atomic-level descriptors are defined as \cite{doi:10.1021/acs.jpcc.9b03925}:

\begin{equation}
    \textbf{v}_{i,\alpha ; k_{[i,j]}}^{[\psi_{i},\psi_{j}]} = \frac{1}{a_{k_{[i,j]}}^{3}(2\pi)^{3/2}} \sum _{j \ne i} \frac{r_{ij}^{\alpha}}{r_{ij}} exp \left (-\frac{(r_{ij} - a_{k_{[i,j]}})^{2}}{w^{2}}  \right ) f_{cut}(r_{ij})
\end{equation}

Here, $r_{i}$ and $r_{j}$ are the Cartesian coordinates of atoms $i$ and  $j$, $r_{ij} = |$\textbf{r}$_{j}$ - \textbf{r}$_{i}|$, $r^{\alpha}_{ij}$ is the projection of \textbf{r}$_{j}$ - \textbf{r}$_{i}$ onto any arbitrary direction $\alpha$.  $[\psi_{i},\psi_{j}]$ represents the chemical identities of the interaction. $k_{[i,j]}$ represents a given atomic feature for the specific $[i,j]$ interaction. The summation runs over the neighbor list set $\{ j \}$ of atom $i$, while $\frac{1}{k_{[i,j]}^{3}(2\pi)^{3/2}}$ is a normalization constant. The damping function $f_{cut}(r_{ij}) = \frac{1}{2}[cos(\frac{\pi r_{ij^{2}}}{R_{cut}^{2}} ) + 1]$, which accounts for a smooth degradation in an atom's contribution to the atomic force exerted on atom $i$, has a cut-off radius $R_{cut}$ chosen to be 7 {\AA}. The gaussian functions are placed at various distances, $a_{k_{[i,j]}}$ away from atom $i$, with widths controlled by $w$. The $a_{k_{[i,j]}}$ values are controlled manually by matching their peaks with those observed in the radial distribution function of the material. Once the peaks have been matched, the remaining distances are then filled in by placing equidistantly spaced features. The gaussian functions are placed from 0.1 {\AA} to 7 {\AA}, to ensure that the underlying potential energy surface is appropriately sampled. In this work, the final atomic feature vector, for a given atom $i$, along direction $\alpha$ will be given as:

\begin{equation}
	V_{i,\alpha} =\{ {v}_{i,\alpha ; k_{[i,i]}}^{[\psi_{i},\psi_{i}]} , {v}_{i,\alpha ; k_{[i,j]}}^{[\psi_{i},\psi_{j}]} \}
\end{equation}

\subsection{Model Construction}

\begin{figure*}
       \centering
    	\includegraphics[width=1.0\textwidth]{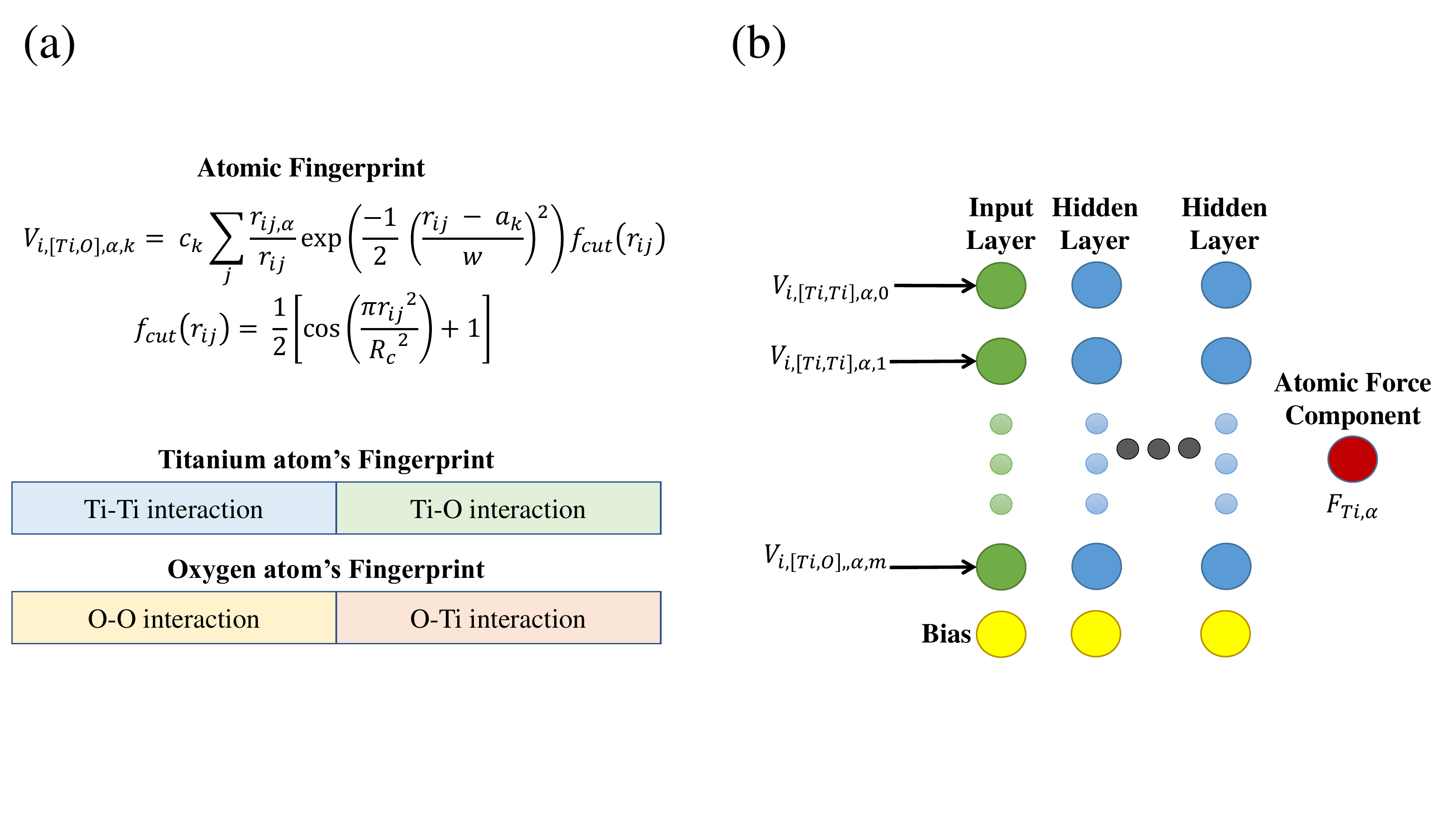}
    	\caption{(a) Workflow of the atomic feature construction process for the titanium AFNN. The atomic feature functional form and corresponding eometric damping function (top) for the case of a titanium atom are shown. A visual depiction of the concatenated final feature vector for titanium and oxygen (bottom) are provided. The current specie's self interaction features are always included first, while the corresponding hetergenous interactions are included after the self-tineraction terms. (b) Visual description of the AFNN architecure is shown. The entire concatenated feature vector is fed into the AFNN, along with a bias term. This information is passed through several hidden layers, culminating in the prediction of a particular atomic force component. The example shown here is for the case of titaium.}
    \label{fig:ffw}
\end{figure*}

The Atomic force neural network (AFNN) employs a feature set containing self-interaction and hetergenous interaction terms \cite{doi:10.1021/acs.jpcc.9b03925}. Each term includes 48 gaussian functions, with means spaced 0.14375 $\AA$ apart as they move away from an atom's center. The first gaussian is palced 0.1 $\AA$ away from an atom's center. The width of each gaussian was set uniformly at 0.2 $\AA$. The same feature set was used for both chemical species studied in this work. The final feature vector contained 96 features, as it is a concatenation of interaction terms. Figure \ref{fig:ffw} provides a visual depiction of the functional form used to calculate the feature vectors, as well as the interaction term concatenation process.

Table \ref{tab:ffdetails} provides technical information regarding the neural network artchitectures used for the Ti and O AFNNs. Figure \ref{fig:ffw} (b) also provides a visual description of the AFNN setup. For both Ti and O AFNNs, 96 features were fed into the input layer of the neural network, along witha  single bias term. The input layer was connected to a hidden layer containing 192 neurons and a bias node. A second, identical hidden layer was connected to the second layer, followed by an final expanded layer containing 768 neurons. These 768 neurons were summed together linearly to predict the atomic force component. The artchitecture employed here was identical between the chemical specie's respective models.

\clearpage

\begin{table}
\centering
    \caption{Neural network architecture for the oxygen and titanium AFNNs. The number of layers provided does not include the input layer.}
\resizebox{10cm}{!}{
	    \begin{adjustbox}{width=\textwidth}
		    \begin{tabular}{c c c}
    			\hline
    			Chemical Identity & \# Layers & \# Neurons  \\
    			\hline
    			Ti   & 3 & 192, 192, 768 \\
			O   & 3 & 192, 192, 768 \\

    			\hline
		    \end{tabular}
    	\end{adjustbox}
}
\label{tab:ffdetails}
\end{table}

The final mathematical form of the NN is expressed in matrix form as:

\begin{equation}
    \vv{u}^{(1)}(i) = \vv{V}_{i}*\hat w^{(0,1)}
\end{equation}
\begin{equation}
    \vv{u}^{(n)}(i) = \vv{f}(\vv{u}^{(n-1)}(i))*\hat w^{(n-1,n)} + b^{(n-1,n)} \ni n > 1
\end{equation}

Here, $\vv{u}^{(n)}$ represents a given set of neurons for a particular layer $n$, with $n=1$ representing the input layer. $\vv{V}_{i}$ represents the atomic feature vector used for a particular atom $i$. $w^{(n-1,n)}$ is the weight matrix of size $(M_{n-1}$ x $M_{n})$, where $M_{n}$ is the number of neurons is a given layer $n$. As mentioned previously, $\vv{f} = \frac{e^{x} - e^{-x}}{e^{x} + e^{-x}}$. $ b^{(n-1,n)}$ represents the bias term associated with a given neuron, outside of the input layer. The final layer used to predict a given atomic force component, is given as $f_{i} = b^{last} + \sum_{m}^{M_{last}}u_{m}^{last}(i)$.

\subsection{Prediction of atomic forces}

\begin{figure*}
       \centering
    	\includegraphics[width=1.0\textwidth]{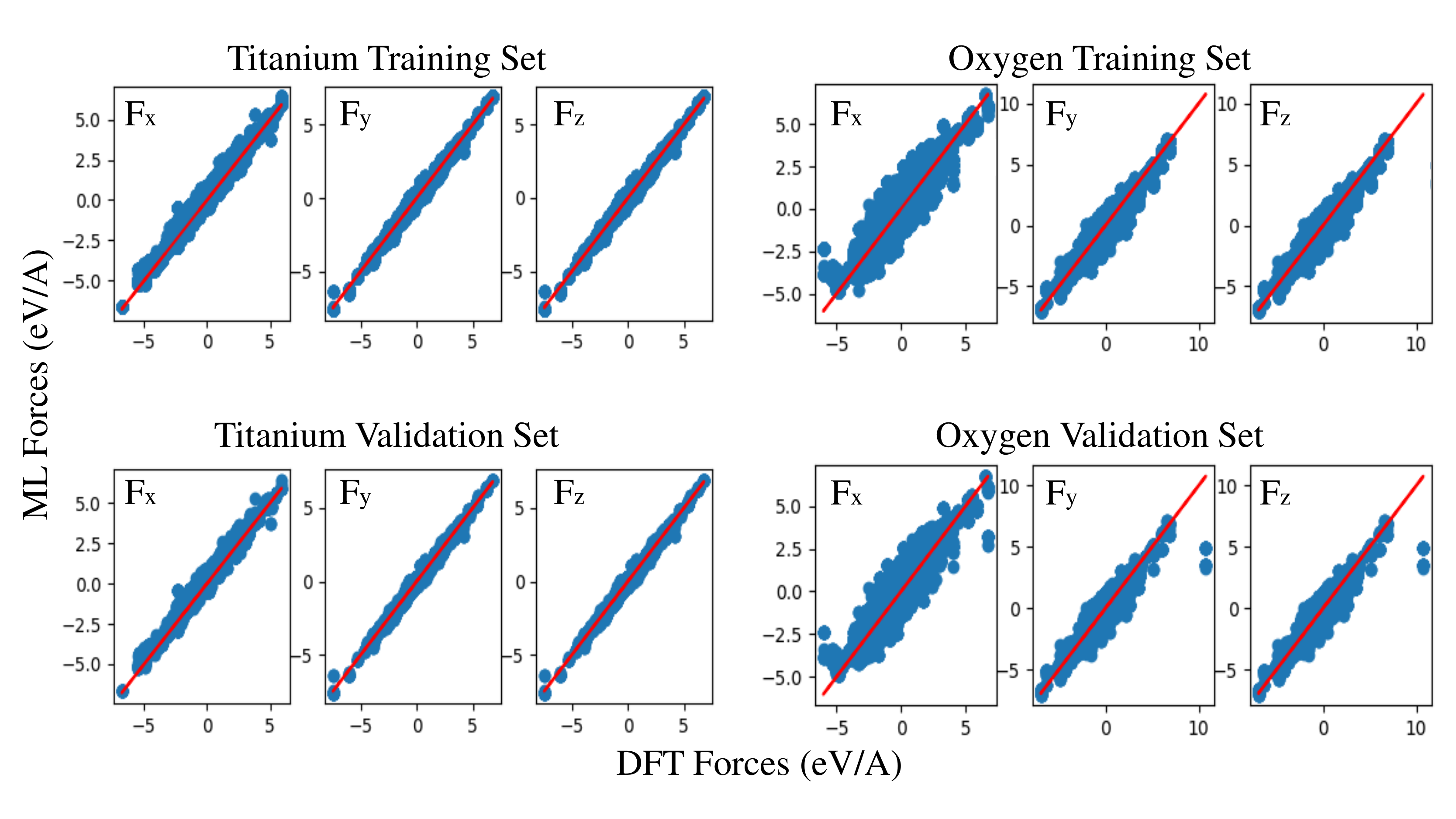}
    	\caption{(left) Atomic forces for the training (top) and validation (bottom) sets for titanium force prediction, shown against the corresponding DFT predictions. (right) Atomic forces for the training (top) and validation (bottom) sets for oxygen force prediction, shown against the corresponding DFT predictions.}
    \label{fig:fp}
\end{figure*}

Figure \ref{fig:fp} shows parity plots for the AFNN and density functional theory (DFT) predicted atomic forces for the DFT-generated reference data. The top plots show the force predictions on the training sets of Ti and O, while the bottom plots provide forces for the validation sets. Good overall agreement is shown for both species, though a larger spread exists for the case of oxygen. Detailed statistics used to validate the AFNN can be found in Table \ref{tab:ffstats}

\begin{table}
\centering
    \caption{Statistics for the atomic force preditions, suing the AFNN, compared to DFT. RMSE indicates the root mean square error, STD is the standard deviation, and KS represents the the p-value from a Kolmogorov-Smirnov \cite{doi:10.1080/01621459.1951.10500769} test.}
\resizebox{10cm}{!}{
	    \begin{adjustbox}{width=0.8\textwidth}
		    \begin{tabular}{c c c c c}
    			\hline
    			Element & Data Set & RMSE ($\frac{eV}{\AA^{2}}$) & STD ($\frac{eV}{\AA^{2}}$) & KS \\
    			\hline
    			Ti  & & & & \\
			& Training   & 0.15  & 0.16 & 0.97 \\
			& Validation   & 0.17 & 0.17 & 0.95 \\
			O   & & & \\
			& Training   & 0.34 & 0.32 & 0.91 \\
			& Validation   & 0.38 & 0.38 & 0.9 \\

    			\hline
		    \end{tabular}
    	\end{adjustbox}
    	\label{tab:ffstats}
}
\end{table}

\subsection{Dynamic properties of TiO$_{2}$}

Mean square displacements (MSD) were calculated using classical molecular dynamics (MD) and ab intio MD (AIMD). Classical MD simulations were performed in LAMMPS \cite{lammps} using the AFNN. Each chmical species was tracked over the course of the simulation. Time was broken into windows (shifted initial temporal starting points) to allow for a smoother resulting MSD. A fickian diffusion model \cite{C5NJ02836A} was used to extract the self-diffusion constants from the resulting MSDs. Figure \ref{tab:MSD} shows the MSDs and corresponding self-diffusion constants for both Ti and O, showing good agreement between DFT and the AFNN.

\begin{figure*}
       \centering
    	\includegraphics[width=1.0\textwidth]{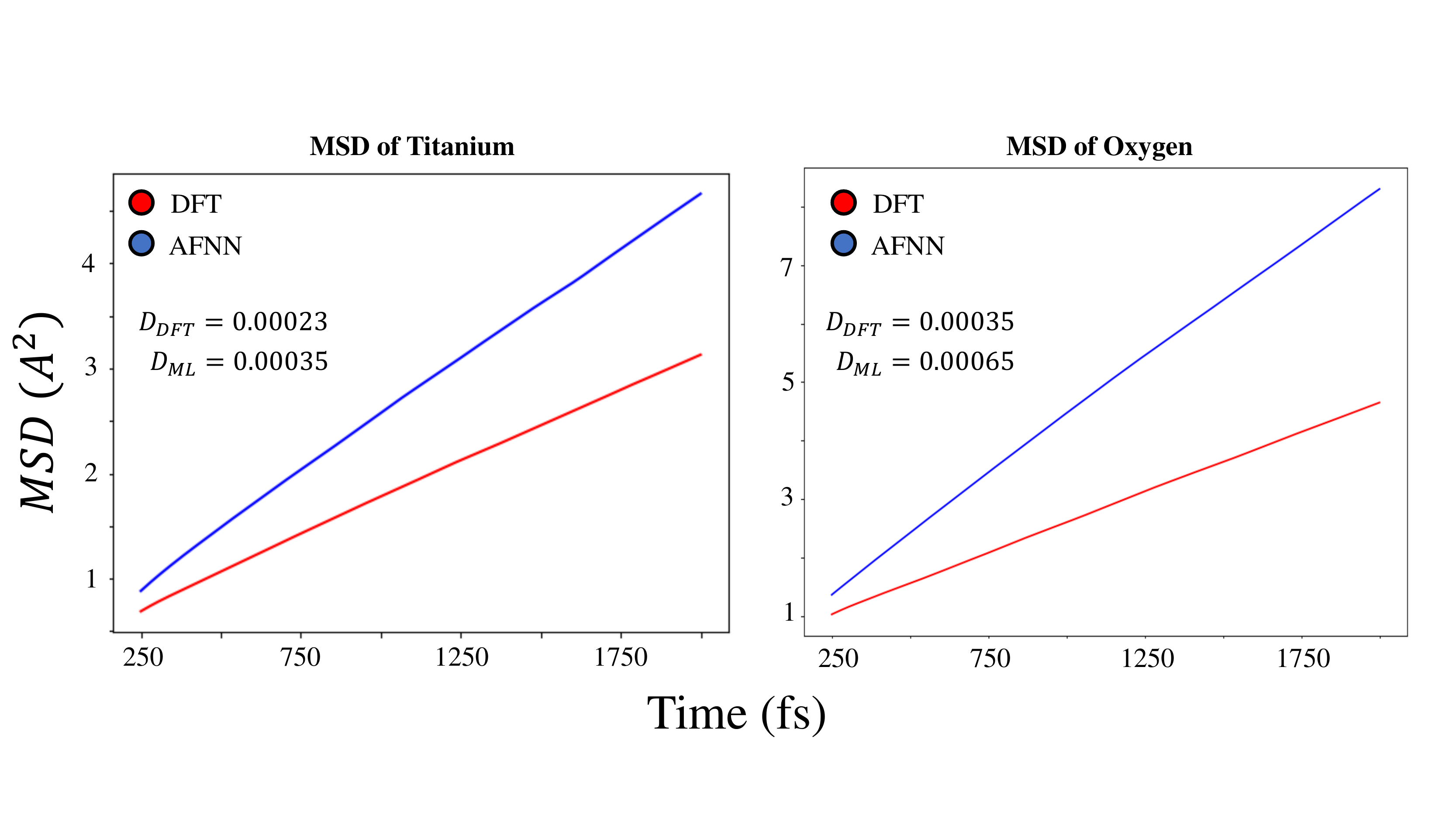}
    	\caption{(left) Mean square displacement of titanium atoms at 2550K, plotted against time, shown in femtoseconds, for AFNN MD (blue) and AIMD (red) simulations. (right) Mean square displacement of oxygen atoms at 2550K, plotted against time, shown in femtoseconds, for AFNN MD (blue) and AIMD (red) simulations.  Inserted values indicate the corresponding diffusion constant, calculated from the slope of the MSD.}
    \label{fig:MSD}
\end{figure*}

\subsection{Structural properties of non-stoichiometric TiO$_{1.88}$}

\begin{figure*}
       \centering
    	\includegraphics[width=0.75\textwidth]{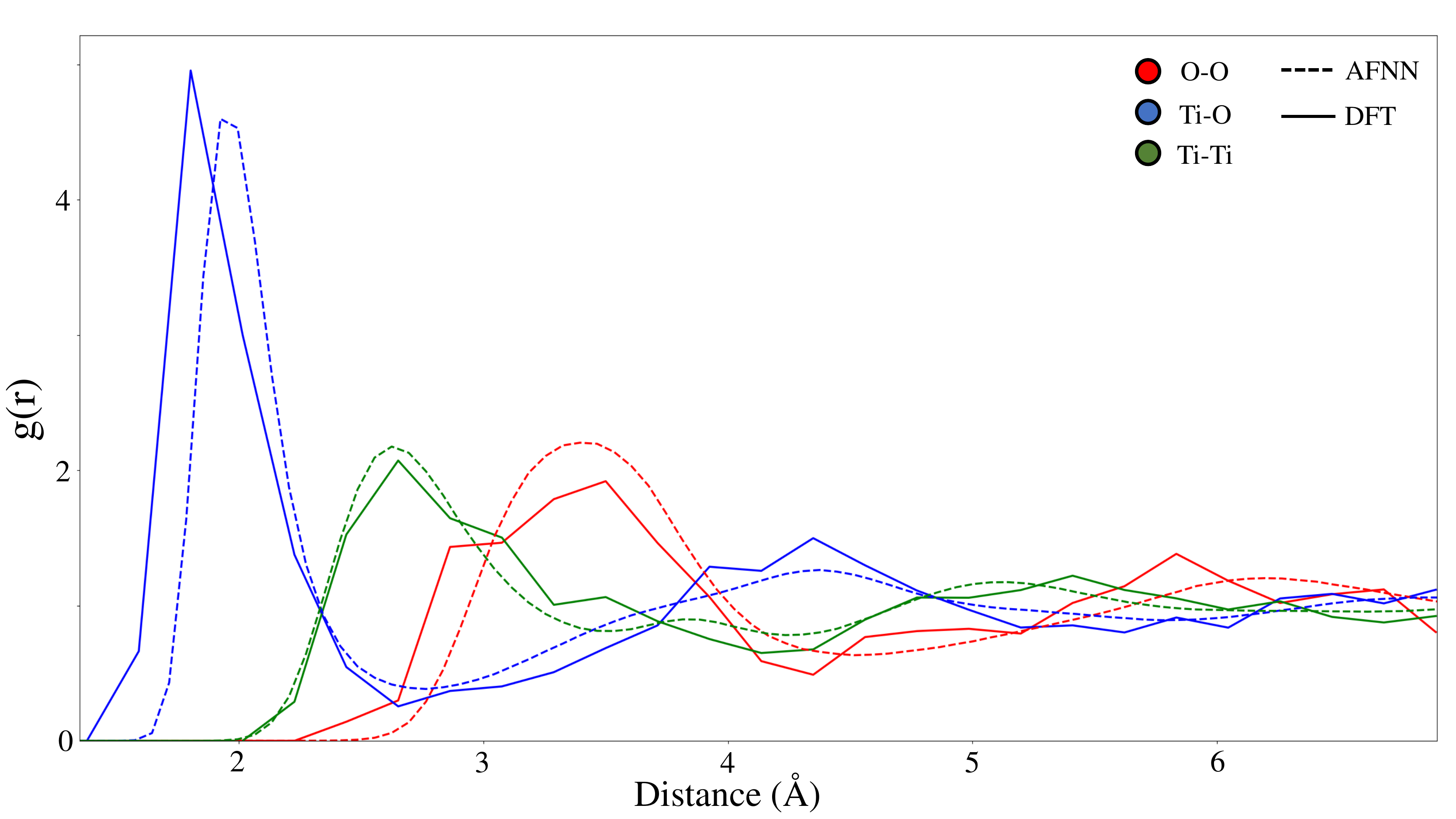}
    	\caption{ Molecular dynamics derived time-averaged pair-correlation function of amorphous TiO$_{1.88}$ at T=2250K, for both DFT and the AFNN. The pair-correlation function is decomposed based on chemical species interactions (colors). The AFNN is shown as the solid line, while DFT is given as a dashed line.}
    \label{fig:188rdf}
\end{figure*}

Time-averaged radial distribution function (RDF) for the AFNN, and static RDF for DFT, were calculated to show the structural agreement between the AFNN and DFT for the case of TiO$_{1.88}$. Results can be found in Fig. \ref{fig:188rdf}. The AFNN predicts a slight shift to the right in the first peak of the RDF, correpsonding to Ti-O interactions. This shift is primarily due to a small shift to the left of the O-O peak, indicating that oxygen atoms are slightly closer together in the AFNN systems. However, it should be noted that the AFNN simulations cover a much larger region of the TiO$_{1.88}$ phase space than DFT, which only contain a single snapshot. Figure \ref{fig:188rdf} shows the RDF predictions for the AFNN and DFT,decomposed into O-O, Ti-O, and Ti-Ti interaction curves.

\subsection{Graph characterization of amorphous phase space}

\begin{figure}
       \centering
    	\includegraphics[width=0.85\textwidth]{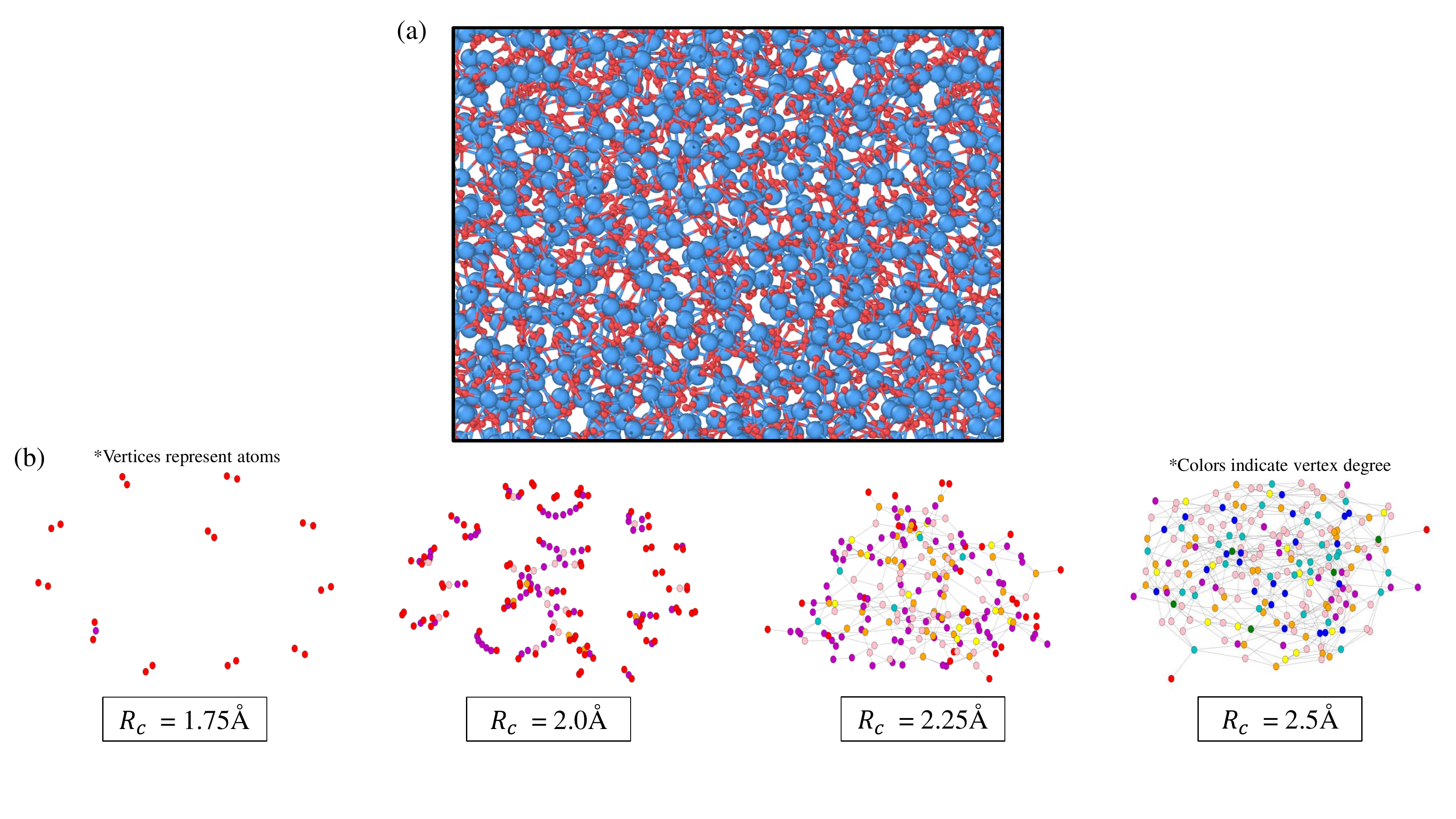}
    	\caption{(a) Atomic-level structure of amorphous TiO$_{2}$ with red atoms representing oxygen and blue atoms representing titanium. (b) Graph coordination networks of (a) at various cutoff distances. Colors are relative to the smallest vertex degree within the graph, with purple representing the smallest degree and blue representing the largest degree. One can see that as the graph cutoff is increased, the GCN becomes increasingly interconnected and chaotic, providing a good indication of the local environments at various cutoff radii.}
    \label{fig:VGOP}
\end{figure}

The GCN is described by an adjacency matrix, with matrix elements defined as:

\begin{equation}
	G_{k_{i},k_{j}}^{i-j} = \frac{1}{d_{k_{i},k_{j}}}  \ni d_{k_{i},k_{j}} \leq R_{c}
\label{equ:GCN}
\end{equation}

Here, $i$ and $j$ represent the chemical species of the atoms contained in the GCN. $k_{i}$ and $k_{j}$ are the atomic indices of a particular atom, from chemical specie $i$ and $j$ respectively. $d_{k_{i},k_{j}}$ is defined as the $l^{2}$-norm between two atoms. $R_{c}$ is the cutoff radius specified when constructing the GCN. Each matrix element, $\frac{1}{d_{k_{i},k_{j}}}$, represents the weight of a given edge for a specific pair of adjacent nodes in the graph. The degree of each node is defined as the sum of the elements in a node's edge set. The degree sets are then fed into the scalar graph order parameter (SGOP) \cite{chapman2021sgop} scheme for the final characterization of the atomic environments. In this work Ti-O GCNs are used to classify the amorphous titania phase space. The SGOP functional form is defined as:

\begin{equation}
    \theta_{i-j,R_{c}} =  \sum_{s}^{S}\left( \sum_{m}^{D_{s}}P(d_{m})\log_{b}P(d_{m}) + d_{m}P(d_{m})  \right)^{3}
\end{equation}

Here, $i$ and $j$ represent the chemical identities of the atoms contained in the GCN. $R_{c}$ is the cutoff radius specified when constructing the GCN. We make the assumption that a particular GCN is disconnected, and that the underlying network exists as a set of subgraphs, $S$, with $s$ indexing a particular subgraph. Note that in the event a GCN is fully connected the outer sum disappears and no further changes are required to the formalism. $D_{s}$ is the set of unique node degrees in a subgraph, with $P_{d_{m}}$ being the probability of a given degree, $d_{m}$, occurring in the subgraph. A Vector Graph Order Parameter (VGOP) \cite{chapman2021sgop} is then assembled from a list of SGOP values, calculated using a cutoff radius set of ${1.75 \AA , 2.0 \AA, 2.25 \AA, 2.5 \AA, 2.75 \AA, 3.0 \AA}$, which were chosen based on the profile of the first Ti-O peak in the radial distribution function. Principal component analysis \cite{karamizadeh_overview_2013} (PCA) is used to reduce the number of features in the VGOP and allow for the visual inspection of the underlying data. Z-score normalization \cite{8667324} is used to normalize the VGOPs to aid in the PCA decomposition. In this work the first two principal components comprised at least 95$\%$ of the underlying variance, and therefore the remaining components were discarded. 

\begin{figure*}
       \centering
    	\includegraphics[width=0.85\textwidth]{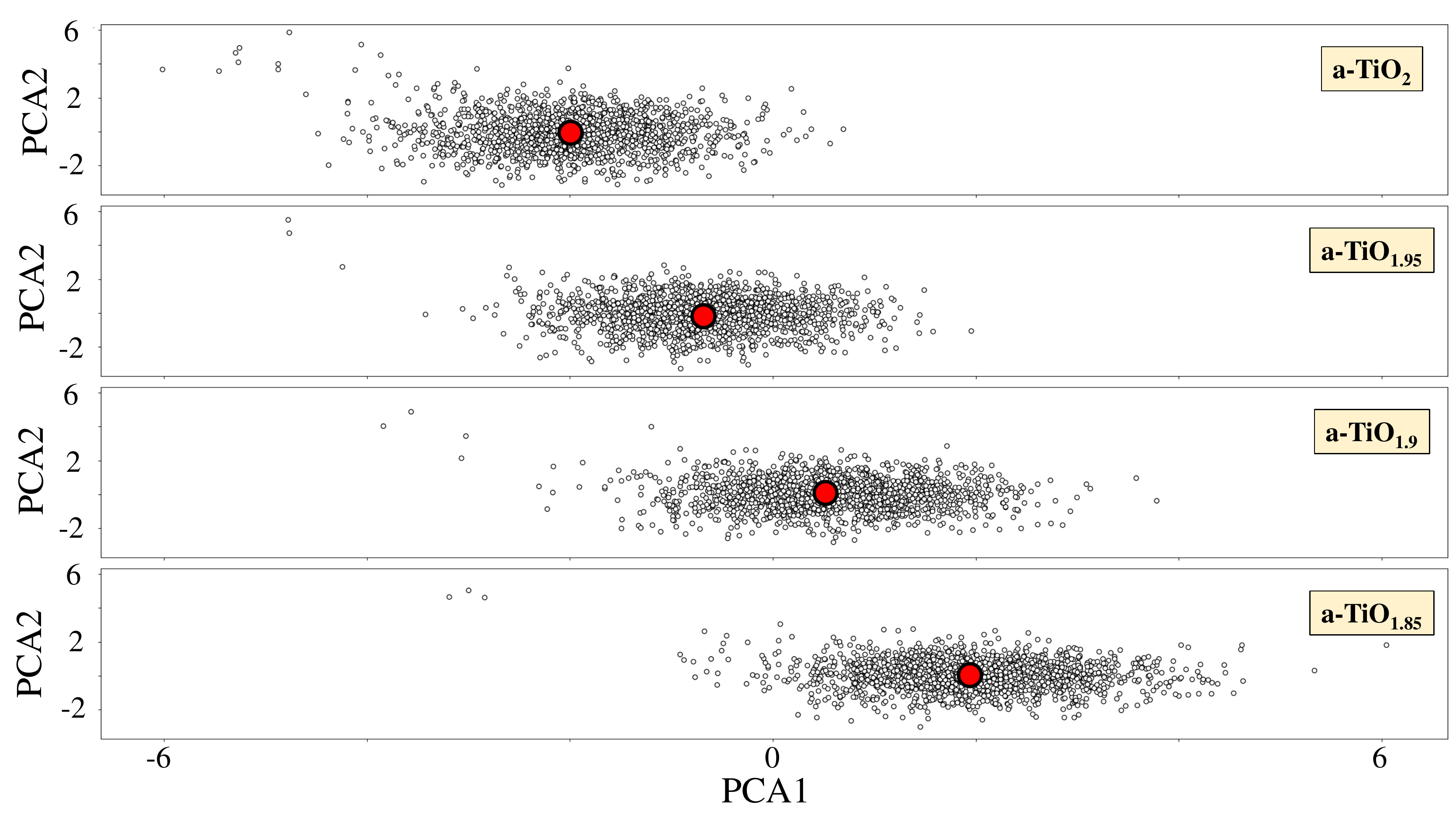}
    	\caption{Decomposed PCA reduction for the amorphous phase space, as characterized using VGOP. Each row corresponds to a unique stoichiometry. Red points indicate the location each cluster's centroid. White points shown here represent approximately 25\% of the total amount of data, which is reduced here for visualization purposes.}
    \label{fig:stacked_pca}
\end{figure*}

As discussed in the main text, the 1944-atom trajectories were used to explore the configuration space of the amorphous phase. Four stoichiometries were considered for the case of TiO$_{x}$ ($x = $ 2.0, 1.95, 1.9, 1.85), to gauge how the structural space changes as a function of oxygen concentration. Figure \ref{fig:stacked_pca} shows the VGOP PCA decomposition for all cases considered in this work. The red points in Figure \ref{fig:stacked_pca} correspond to the centroid of each cluster, which is calculated as the average x and y coordinate over all points in a given stoichiometry. The location of the red points corresponds to the black points shown in Figure 5 (b) in the main text. A subset of the total dataset was used in Figure \ref{fig:stacked_pca} to give the reader a better idea of how the majority of each phase's points radially extend away from the cluster's centroid.

\section{DFT Binding Energies}

\subsection{96-atom unit cell}

\begin{figure}
       \centering
    	\includegraphics[trim={0 0cm 0 0cm},width=1.0\textwidth]{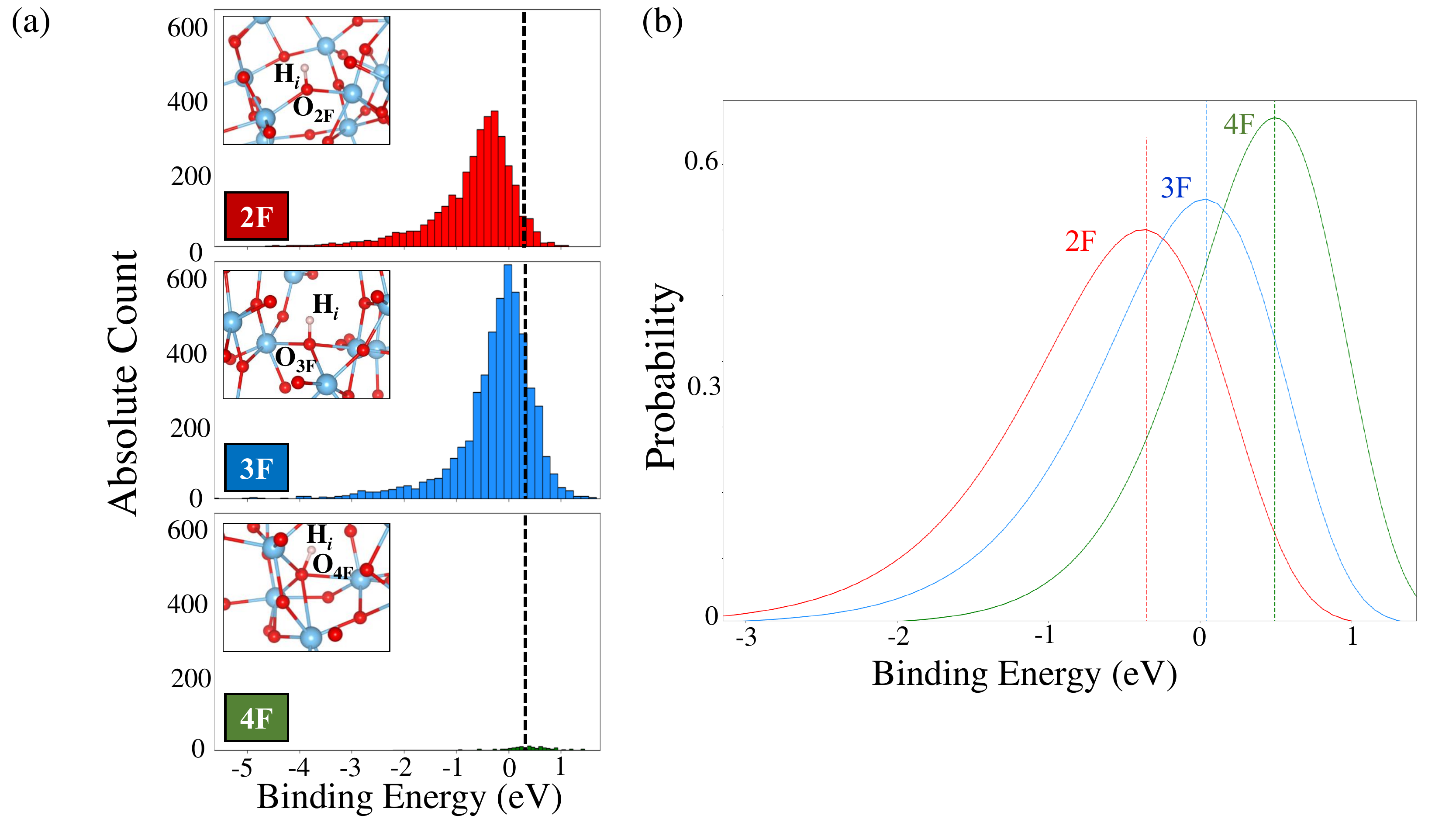}
    	\caption{(a) Histograms of the hydrogen binding energies for the various CN environments. Values shown here are the absolute number of samples, signifying the significant reduction in data set size for 4F environments. Values are color coded based on the CN environment. The dashed vertical lines indicate the mean for the distributions. (b) Fitted hydrogen binding energies for TiO$_{1.88}$. The distributions are colored according to the coordination number of the particular oxygen site. Here, the x-axis represents the binding energy, while the y-axis represents the probability of that binding energy occurring with respect to the number of environments for that CN type. The dashed black line represents the hydrogen binding energy in crystalline rutile. Inserted images in (a) show the oxygen environments encountered by hydrogen in the amorphous TiO$_{1.88}$ phase space.}
    \label{fig:Hbind}
\end{figure}

Binding energy distributions are shown for the non-stoichiometric case of TiO$_{1.88}$ in Fig \ref{fig:Hbind}. Binding energy distributions are nearly identical between TiO$_{1.88}$ and TiO$_{2}$, indicating that the concentration of oxygen in the system plays little-to-no role in the hydrogen insertion energetics, at least for the cases examined in this work. 

\subsection{216-atom unit cell}

\begin{figure}
       \centering
    	\includegraphics[trim={0 0cm 0 0cm},width=1.0\textwidth]{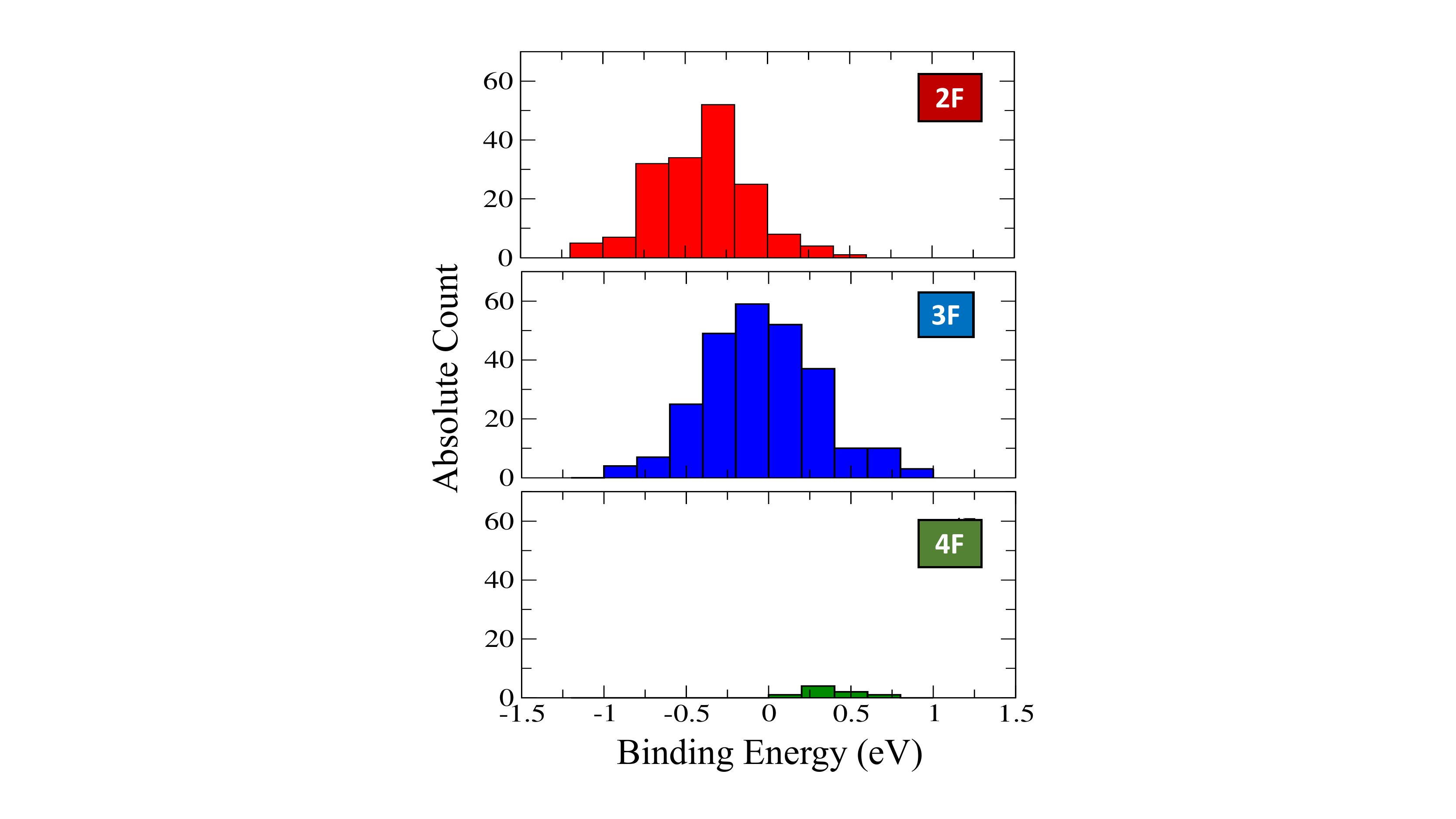}
    	\caption{Histograms of the hydrogen binding energies for the various CN environments. Values shown here are the absolute number of samples, signifying the significant reduction in data set size for 4F environments. Values are color coded based on the CN environment. Here, the x-axis represents the binding energy.}
    \label{fig:216bind}
\end{figure}

Binding energy distributions are shown for the 216-atom unit cell systems of TiO$_{2}$ in Fig \ref{fig:216bind}. 216-atom systems are taken from the DFT training data. While the number of binding energies represented here is significantly less than that of the 96-atom case, due to the cost of performing DFT on 216-atom unit cells, a trend does emerge for the 216-atom case in which binding energies less than -2 eV are not seen, while there exist many such configurations for 96-atom structures. While this may be due to the limited number of samples in the 216-atom case, it may be possible that such an effect is due to the hydrogen concentration, however we do not examine this in detail in this work.

\section{Experimental Characterization}

\begin{figure}
       \centering
    	\includegraphics[width=0.85\textwidth]{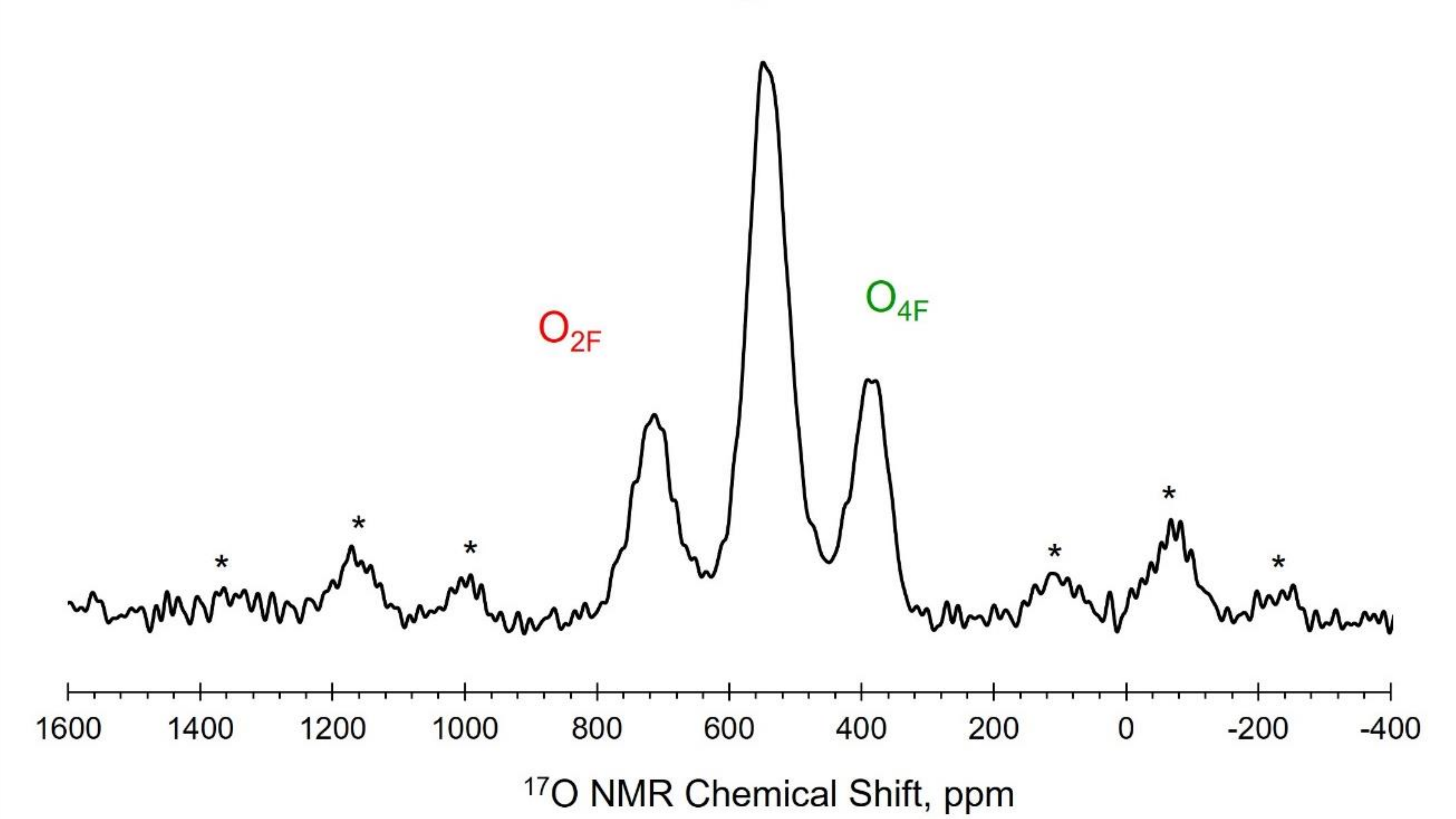}
    	\caption{Experimental $^{17}$O solid-state NMR of synthesized amorphous TiO$_{2}$ powder. The spectrum was referenced to tap water. The asterisks denote the sidebands. The synthesis procedures, phase, and stoichiometry characterizations for the powdered TiO$_{2}$ are given in supplementary information. }
    \label{fig:full_NMR}
\end{figure}

\begin{figure}
       \centering
    	\includegraphics[width=0.85\textwidth]{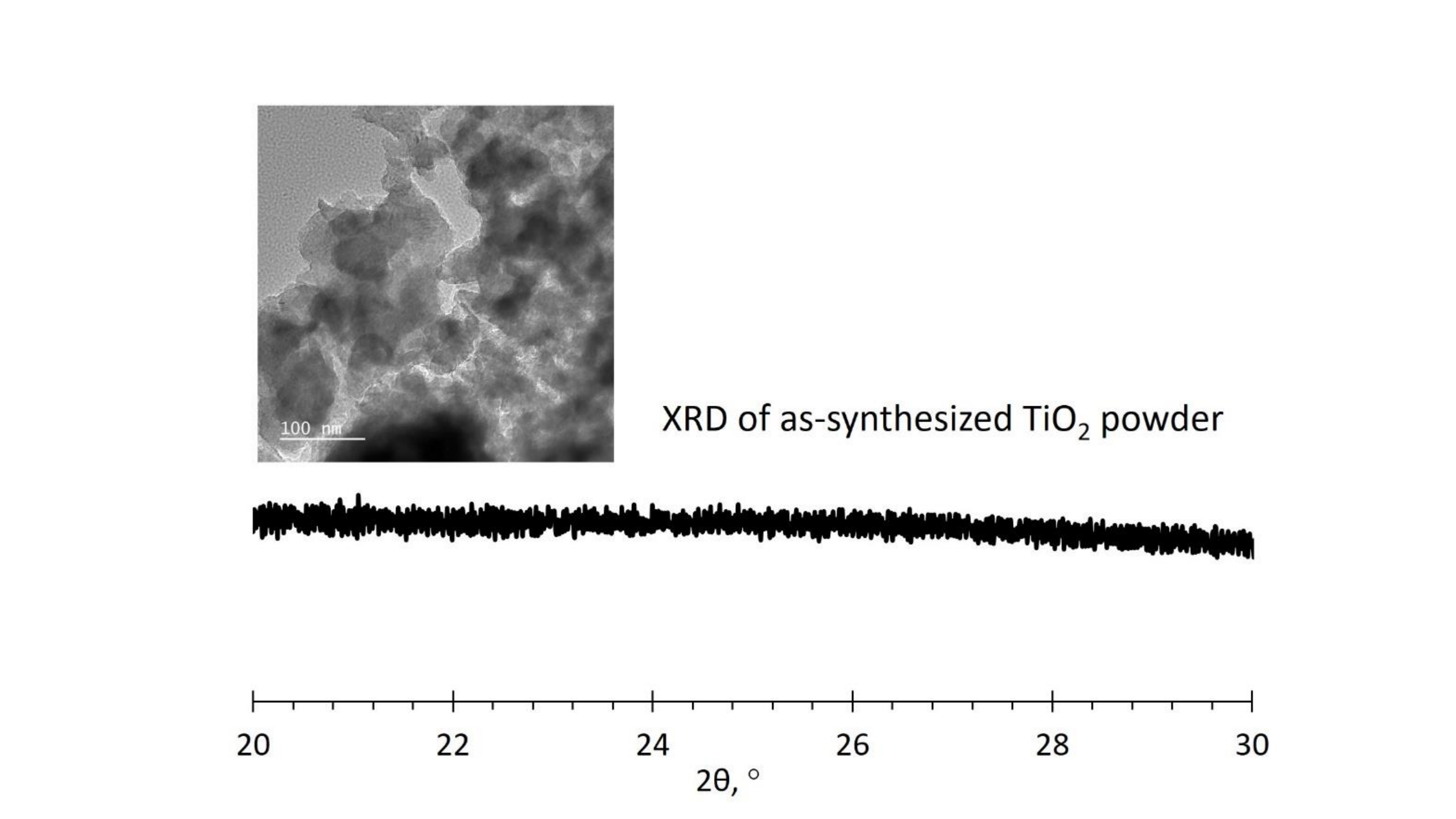}
    	\caption{TEM imaging and XRD spectra of synthesized TiO$_{2}$ powder.  The synthesized TiO$_{2}$ powders aggregates into nanometer scale clusters and are amorphous phase. }
    \label{fig:XRD}
\end{figure}

\begin{figure}
       \centering
    	\includegraphics[width=1.0\textwidth]{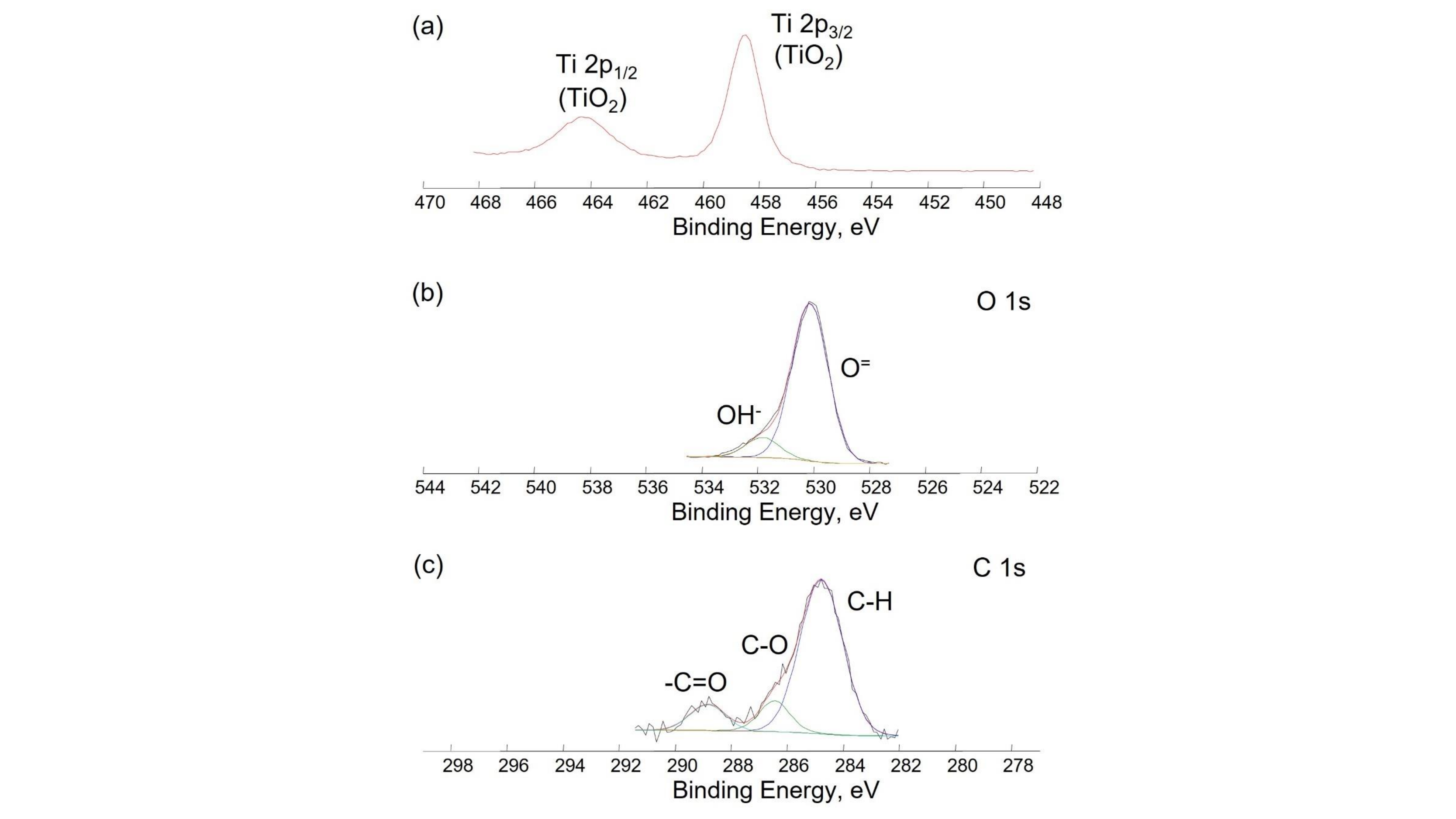}
    	\caption{XPS spectra for the examination of chemical species and stoichiometry: (a) Ti 2p$_{3/2}$ and Ti 2P$_{1/2}$ peaks that corresponding to TiO$_{2}$. The stoichiometry of Ti oxide is 2 and is in align with the RBS results. (b) O 1s peak corresponding to OH- and O= groups. (c) C 1s peak corresponding to -C=O, C-O and C-H groups where were sourced from the precursor for TiO$_{2}$ synthesis. The atomic ratio of O/Ti is about 2.6, which is higher than 2 in TiO$_{2}$. The extra oxygen is contributed by the OH-, C-O, and -C=O groups. }
    \label{fig:XPS}
\end{figure}

\begin{figure}
       \centering
    	\includegraphics[width=0.85\textwidth]{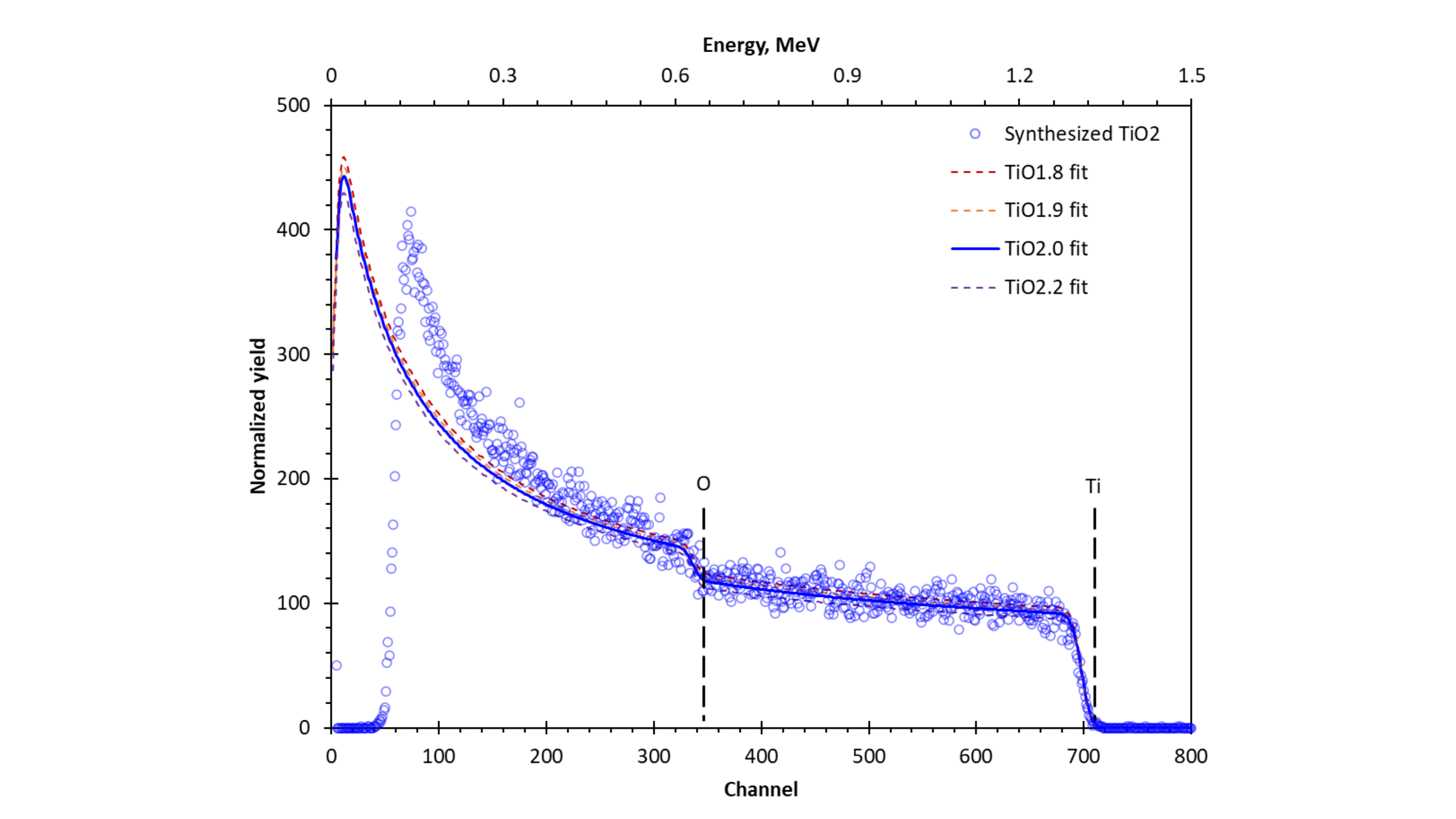}
    	\caption{Rutherford backscattering spectra from TiO$_{x}$ film. Symbols are experimental points, while solid lines are results of RUMP-code simulations. For clarity, only every 15th experimental point is depicted. The position of $^{16}$O, and $^{48}$Ti are peaks are marked by arrows. The best fit to the experimental data is obtained with O/Ti = 2.}
    \label{fig:RBS}
\end{figure}
\section{Conclusion}

The corrosion of materials in a multitude of environments presents a significant economic and technological burden. This work aims to understand the atomic-level precursor processes that lead to the eventual failure of metals and metal alloys by characterizing the structural and thermodynamic properties of hydrogen within amorphous titania. 
We show that, through a combination of simulations of experiments, that we can connect and validate the atomic structure of amorphous titania. We quantify the type of oxygen site based on its corresponding coordination number using NMR simulations, which is validated by experiments. Through molecular dynamics and graph theory we see that the amount of various O CN environments are stoichiometry-dependant, which is used to validate the experimental characterization using NMR. 

We show though DFT calculations, that hydrogen preferentially binds to 2F coordinate oxygen atoms, across several stoichiometries, due to its strong binding energy. We also show that each O CN environment exhibits a spectrum of H binding energies. The H binding energies, combined with the stoichiometry-dependant rations of the various O CN environments, paints a picture where H diffusion can be affected by both the local oxygen energetics as well as the spatial properties of long-range CN networks. If correct, this implies that hydrogen diffusion in amorphous titania could be controlled simply by tailoring the amount of oxygen present within the system. The tools and analysis presented in this work provide a simple and straightforward pathway to potentially understand the atomistic mechanisms behind properties such as incubation time, permeation, and ultimately corrosive failure of the underlying material.

\section{Supporting Information}
This work contains supplemental information which can be found online.

\section{Data Availability}
The TiO$_{x}$ AFNN created and used in this work will be included as part of the LAMMPS distribution.

\section*{Acknowledgements}
J. Chapman, K. E. Kweon, N. Goldman, N. Keilbart, Y. Zhu, T. W. Heo, and B. Wood are partially supported by the Laboratory Directed Research and Development (LDRD) program (20-SI-004) at Lawrence Livermore National Laboratory. This work was performed under the auspices of the US Department of Energy by Lawrence Livermore National Laboratory under contract No. DE-AC52-07NA27344. K.B. acknowledges that this material is based upon work supported by the U.S. Department of Energy, Office of Science, Office of Advanced Scientific Computing Research, Department of Energy Computational Science Graduate Fellowship under Award Number DE-SC0020347. J. Chapman gratefully acknowledges the support of the College of Engineering and Department of Mechanical Engineering
at Boston University. 
 
\section*{Author Contributions}

J. Chapman constructed the AFNN model and performed all AFNN simulations. K. Kweon generated DFT-MD data. K. Kweon performed all DFT binding energy calculations. K. Kweon and K. Bushick performed all DFT-NMR calculations. J. Chapman performed all graph characterization calculations. J. Chapman created and executed the automated hydrogen insertion pipeline. Y. Zhu performed all experimental measurements with the help of R. Quiu, L. Aji, and C. Colla. J. Chapman, K. Kweon, Y. Zhu, and B. Wood wrote the manuscript with inputs from all authors. N. Goldman and B. Wood supervised the research.

\section*{Competing Interests}
The authors declare no competing financial or non-financial interests.

\printbibliography

\end{document}